**[Title]** A fast and integrative algorithm for clustering performance evaluation in author name disambiguation

**[Authors]** Jinseok Kim

**[Author Information]**

Jinseok Kim

Institute for Research on Innovation & Science, Survey Research Center, Institute for Social Research, University of Michigan
330 Packard Street, Ann Arbor, MI U.S.A. 48104-2910
734-763-4994|jinseokk@umich.edu

Abstract

Author name disambiguation results are often evaluated by measures such as Cluster-F, K-metric, Pairwise-F, Splitting & Lumping Error, and B-cubed. Although these measures have distinctive evaluation schemes, this paper shows that they can be calculated in a single framework by a set of common steps that compare truth and predicted clusters through two hash tables recording information about name instances with their predicted cluster indices and frequencies of those indices per truth cluster. This integrative calculation reduces greatly calculation runtime, which is scalable to a clustering task involving millions of name instances within a few seconds. During the integration process, B-cubed and K-metric are shown to produce the same precision and recall scores. In this framework, especially, name instance pairs for Pairwise-F are counted using a heuristic, surpassing a state-of-the-art algorithm in speedy calculation. Details of the integrative calculation are described with examples and pseudo-code to assist scholars to implement each measure easily and validate the correctness of implementation. The integrative calculation will help scholars compare similarities and differences of multiple measures before they select ones that characterize best the clustering performances of their disambiguation methods.

Keywords: author name disambiguation; entity resolution; clustering; evaluation measure; pairwise-F

Introduction

Author name disambiguation is an entity resolution task to generate clusters of name instances to refer to distinct authors in bibliographic data. It is crucial to research that mines authorship data because ambiguous names can lead to merging and/or splitting of author identities and thus flawed knowledge about research production and collaboration (Fegley & Torvik, 2013; Kim & Diesner, 2015, 2016; Strotmann & Zhao, 2012). As publications and ambiguous author names such as East Asian names increase in digital libraries (Bornmann & Mutz, 2015; Torvik & Smalheiser, 2009), various methods for disambiguating author names (Hussain & Asghar, 2017; Smalheiser & Torvik, 2009) have been proposed.

After a disambiguation method is implemented, its clustering result is evaluated by a variety of measures. As there is no consensus on a definitive measure for author name disambiguation (Ferreira, Gonçalves, & Laender, 2012), one or two measures are chosen at the researcher's discretion. The selection of a measure is, sometimes, justified by the argument that it is frequently used or enables the comparison of a study with prior work. In many studies, however, a measure is selected without such clarification.

The clustering measure selection should be understood in the context of each study. It can, however, change our impression about a disambiguation method if its performance is evaluated high by one measure but low or mediocre by another. Applying diverse measures to a disambiguation study can be a non-trivial task because clustering measures have distinct evaluation schemes which are not easy to compare their similarities and differences. In addition, the straightforward implementation of a measure such as Pairwise-F can consume too much runtime depending on data size because the number of instance pairs for comparison can increase quadratically in a worst-case scenario (Menestrina, Whang, & Garcia-Molina, 2010).

To aid scholars to select measures that characterize best their disambiguation results, this study shows that five commonly used measures for evaluating clustering results in author name disambiguation can be calculated all-in-one by implementing a common code. This integrative calculation shows intuitively where those measures are similar and different in evaluating clustering performance. Especially, the proposed approach reduces computation runtime, dramatically for Pairwise-F in particular. In the following sections, the usage patterns of clustering measures in author name disambiguation research are reviewed. Then, the integration process is explained step-by-step with pseudo-code and examples.

Literature Review

Table 1 shows the list of selected author name disambiguation studies and their measures for evaluating clustering performance. Note that detailed explanation of each measure will be provided in the Results section later in this paper.

*Table 1: Clustering performance measures in selected author name disambiguation studies*

| Studies | Cluster-F | K-metric | SE & LE | Pairwise-F | $B^3$ |
|---|---|---|---|---|---|
| Cota et al. (2010) |  | √ |  | √ |  |
| Fan et al. (2011) |  |  |  | √ |  |
| Ferreira et al. (2014) |  | √ |  | √ |  |
| Han et al. (2017) |  |  |  |  | √ |
| Huang et al. (2006) | √ |  |  | √ |  |
| Hussain and Asghar (2018) | √ | √ |  | √ |  |
| Kim and Diesner (2016) | √ | √ |  |  |  |
| Kim and Kim (2018) |  |  |  |  | √ |

| Study | | | | | |
|---|---|---|---|---|---|
| Lerchenmueller and Sorenson (2016) | | | √ | | |
| Levin et al. (2012) | | | | √ | √ |
| Liu et al. (2014) | | | √ | √ | |
| Liu et al. (2015) | | | | √ | |
| Louppe et al. (2016) | | | | √ | √ |
| Momeni and Mayr (2016) | | | | | √ |
| Müller et al. (2017) | | | | | √ |
| Pereira et al. (2009) | √ | √ | | √ | |
| Santana et al. (2017) | | √ | | √ | |
| Shin et al. (2014) | √ | √ | | √ | |
| Qian et al. (2015) | | | | | √ |
| Torvik and Smalheiser (2009) | | | √ | | |
| Wu et al. (2014) | | √ | | √ | |
| Zhang et al. (2018) | | | | √ | |
| Zhu et al. (2018) | | | | √ | |

According to the table, Pairwise-F is the most popular. It appears in 15 out of 23 studies. This confirms that it is the most frequently used in entity resolution in general (Menestrina et al., 2010) as well as in author name disambiguation (Levin, Krawczyk, Bethard, & Jurafsky, 2012)[1]. K-metric is found in 8 studies, followed by B-cubed ($B^3$, 7) and Cluster-F (5). Three studies use the Splitting & Lumping Errors (SE & LE) measure.

In Table 1, 11 out of 23 studies rely on a single measure while others on two or three measures. In addition, the combinations of co-used measures vary. Figure 1 shows the pairs of co-used measures in the Table 1 studies and their co-usage frequencies. For example, Pairwise-F is paired with K-metric 7 times. Interestingly, some possible pairs have never been calculated together. For example, $B^3$ is paired with Pairwise-F twice but not with K-metric, Cluster-F, and SE & LE.

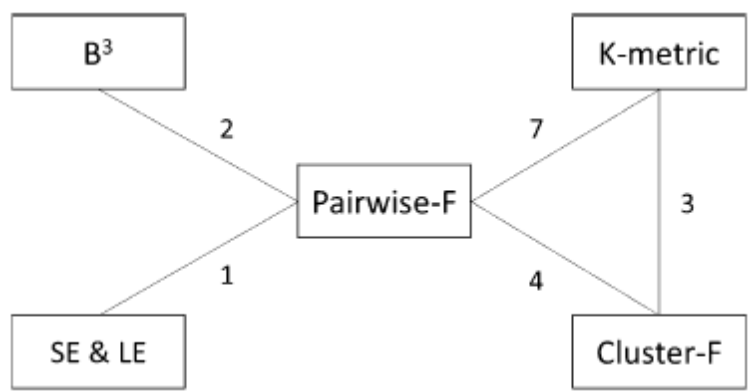

*Figure 1: Co-Usage frequency of pairs of disambiguation measures used together in selected studies in Table 1*

---

[1] Note that B-Cubed is more frequently used than other measures in person name disambiguation on the Web (e.g., Delgado, Martínez, Montalvo, & Fresno, (2017)) because the metric has formal properties that can handle evaluation scenarios specific to the task. For details, see Amigó, Gonzalo, Artiles, and Verdejo (2009).

The use of Pairwise-F is sometimes justified by its frequent usage in entity resolution studies. Other measures are selected to follow the practice of referenced studies to be compared or without any clarification. Although such choices should be understood in each study's unique context, they can change our impression about the clustering performance of a disambiguation method. To illustrate this, the disambiguation performance of a digital library, DBLP (Ley, 2009; Reitz & Hoffmann, 2013), was evaluated on a labeled dataset, KISTI (Kang, Kim, Lee, Jung, & You, 2011). KISTI consists of a set of ambiguous name instances filtered from publication records in DBLP and disambiguated semi-manually by researchers at the Korean Institute for Science and Technology Information. Among 41,673 name instances in the original KISTI, a total of 41,358 name instances are matched to DBLP records[2]. Figure 2 shows the DBLP's clustering performance evaluated on KISTI by five measures.

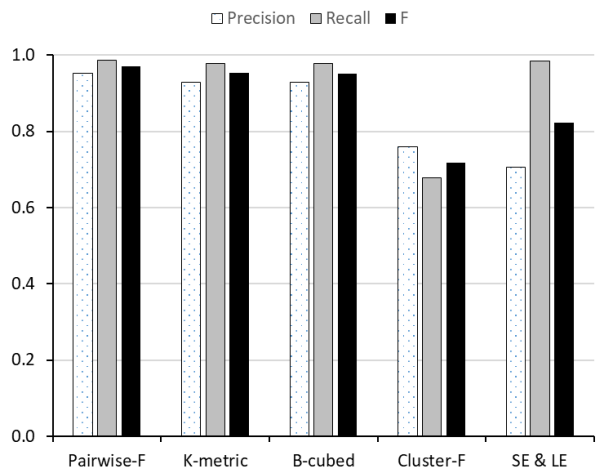

*Figure 2: Performance of DBLP's author name disambiguation evaluated by five measures on KISTI*

Figure 2 shows that DBLP's disambiguation is highly accurate: precision, recall, and F scores of three measures – Pairwise-F, $B^3$, and K-metric – are all above 0.95, corroborating Kim (2018). Cluster-F and SE & LE scores are, however, not so much encouraging. Especially, Cluster-F shows that DBLP performs a little worse in recall than in precision, which contrasts other three measures reporting that DBLP performs better in recall than in precision. According to SE & LE, DBLP disambiguates better regarding recall than precision but the recall-precision performance gap (|recall – precision| = 0.1794) is much pronounced than those by Pairwise-F, K-metric, and B3 (|recall – precision| = 0.0346 ~ 0.0487).

This illustrates why we need to consider various measures for evaluating a disambiguation method. Depending on the choices of measures, the same clustering results can be evaluated as encouraging or less so. As shown in Table 1, however, the selection of measures do not seem to be guided by any common practice. But this does not imply that scholars need to report evaluation results obtained from all available measures, which is undesirable for efficient scientific communication.

Instead, it should be emphasized that the use of diverse measures can illuminate where a proposed disambiguation method performs well and worse. For example, the low Cluster-F coupled with high $B^3$ in Figure 2 indicates that misidentified name instances by DBLP are not many (high $B^3$ scores) but appear across several truth clusters because a single misidentified instance in a truth cluster decides the DBLP's performance for the cluster as a failure. In addition, diverse measures can enable scholars to compare performances of their proposed methods with other studies evaluated by different measures and thus to find room for improvement or synthesize strengths of each study.

---

[2] For details on the matching procedure, see Kim (2018).

Applying different measures to a disambiguation study can, however, be a non-trivial task. Although each measure is well defined in equations, its implementation requires a careful validation of evaluation accuracy. In addition, each measure can be implemented by different code snippets which are not often shared. So, scholars who want to implement a clustering measure often need to write code from scratch. Sometimes, the calculation of a measure such as Pairwise-F may not be easily implementable for a large dataset: it can consume much computing time and RAM because the number of instance pairs can increase quadratically "in the worst case" (Menestrina et al., 2010)[3].

To facilitate the efficient use of diverse clustering measures for author name disambiguation, this study proposes algorithms to calculate the five commonly used measures all-in-one in an integrative framework. Specifically, although the five measures have different evaluation schemes, they can be calculated by implementing a common code, which will help us understand better the similarities and differences of those measures. This integrative calculation is the first attempt of this sort and a novel contribution to the measurement of clustering performance in author name disambiguation. Moreover, during the integration process, $B^3$ and K-metric are shown to produce the same precision and recall scores. Within this framework, especially, Pairwise-F is calculated by a heuristic rather than a brute-force comparison of instance pairs, reducing greatly computation time from quadratic (at worst) to linear one. This solution is motivated by Menestrina et al. (2010) where Pairwise-F is calculated linearly through a 'Slice' algorithm combined with a cost function. This study combines the 'Slice' algorithm with a heuristic to calculate Pairwise-F faster than the 'Slice' algorithm + cost function approach. In following sections, the details of integrative calculation are described with examples and pseudo-code.

## Methods

To evaluate the clustering performance in author name disambiguation, scholars usually measure the similarity between clustering results produced by a disambiguation method and those by human coders in two ways: recall and precision. Here, a cluster consists of name instances that are decided to represent the same authors by an algorithm (a predicted cluster) or manual labeling (a truth cluster). Recall considers how many truth clusters are not compromised by merged or split name instances in predicted clusters, while precision evaluates how many predicted clusters group correctly name instances that belong to the same truth clusters.

Incorporating the aforesaid five measures into the same framework is possible because all of them evaluate performance by both recall and precision. What makes them different is that each measure is designed to assess precision and recall at one of three levels: cluster, instance, or pair of instances, as summarized in Table 2.

*Table 2: Summary of Calculation Level and Recall-Precision Types per Performance Measure*

| **Measure** | **Cluster-F** | **K-metric** | **SE & LE** | **Pairwise-F** | **$B^3$** |
|---|---|---|---|---|---|
| **Calculation Level** | Cluster | Cluster | Cluster | Pair | Instance |

---

[3] For example, a set of 3,964 author name instances can generate over 7.8M instance pairs (Kim, Sefid, & Giles, 2017). To address this challenge in the context of author name disambiguation, a few studies have proposed advanced blocking algorithms. For details, see Kim et al. (2017).

| | | | | | |
|---|---|---|---|---|---|
| **Recall** | Cluster Recall | AAP | Splitting Error | Pairwise Recall | B³ Recall |
| **Precision** | Cluster Precision | ACP | Lumping Error | Pairwise Precision | B³ Precision |
| **F Score** | Harmonic Mean | Geometric Mean | Harmonic Mean | Harmonic Mean | Harmonic Mean |

Despite such different calculation levels, the measures can be implemented by embedding the instance- and pair-level calculations into the cluster level calculation through a set of common code ("skeleton code" hereafter). Algorithm 1 shows the skeleton code.

```
Algorithm 1: Skeleton Code
1      P: a set of predicted clusters
       p: an instance of a cluster P_i
       pIndex: a hash of an instance p and its cluster index i
       T: a set of truth clusters
       t: an instance of a cluster T_j
       tMap: a hash of an instance t and its cluster index i mapped in pIndex
2      pIndex ← {}
3      for each P_i ∈ P do
4          for each p ∈ P_i do
5              pIndex[p] ← i
6          end for
7      end for
8      for each T_j ∈ T do
9          tMap ← {}
10         for each t ∈ T_j do
11             if pIndex[t] ∉ keys(tMap) then
12                 tMap[pIndex[t]] ← 0
13             end if
14             tMap[pIndex[t]] ← tMap[pIndex[t]] + 1
15         end for
16         for each (key, value) ∈ tMap do
17             # do calculation
18         end if
19         end for
20     end for
```

The key idea of Algorithm 1 is that truth clusters are not compared cluster by cluster to predicted ones. Instead, a name instance ($p$) in a predicted cluster ($P_i$) is recorded into a hash table (*pIndex*) where the instance $p$ (key) is mapped to its cluster membership (= $i$: value) (code line #2~#7). Next, a name instance ($t$) in a truth cluster ($T_j$) is checked for its index ($i$) in predicted clusters ($P$) by referencing *pIndex*. Then, the count of the index ($i$) are recorded into another hash table (*tMap*) where an index $i$ (key) is mapped to its frequency (value) (code line #10~#15). In other words, this code snippet counts the number of name instances in a truth cluster that appear together in predicted clusters (= sharing the same $i$), which corresponds to detecting the intersection of a truth cluster ($T_j$) and predicted clusters ($P$). Note that this procedure adopts part of the Slice algorithm in Menestrina et al. (2010).

Within this cluster-level calculation framework, pair- and instance-level measures can be calculated with some modification of their evaluation schemes. To demonstrate this, each measure is explained in detail below starting from cluster-level to pair- and instance-level.

# Results

## Cluster Level: Cluster-F

Cluster-F ($cF$) is a harmonic mean of cluster recall ($cR$) and cluster precision ($cP$) (Menestrina et al., 2010).

$$cR = \frac{|P \cap T|}{|T|} \quad (1)$$

$$cP = \frac{|P \cap T|}{|P|} \quad (2)$$

$$cF = \frac{2 \times cR \times cP}{(cR + cP)} \quad (3)$$

Here, $P$ is a set of predicted clusters, while $T$ is a set of truth clusters. The numerator $|P \cap T|$ counts the number of predicted clusters that contain all and the only instances belonging to the same truth clusters. Cluster recall ($cR$) is the ratio of the numerator over the number of all truth clusters ($|T|$). Cluster precision ($cP$) is the ratio of this numerator over the number of all predicted clusters ($|P|$).

Table 3 shows an example for calculating Cluster-F. In the first column, there are three truth clusters ($T_1$, $T_2$, and $T_3$) in which eight name instances with numeric ids (1, 2, 3...8) are assigned. The second column shows predicted results: eight instances in the first column are assigned to two clusters ($P_1$ and $P_2$). After instances are compared across predicted and truth clusters, only one case of $|P \cap T|$ ($P_1 = T_1$) is detected. So, the numerator for $cR$ is 1, while the denominator is 3 (the number of truth clusters), resulting in $cR = 1/3$. The numerator for $cP$ is also 1 but its denominator is 2 (the number of predicted clusters), resulting in $cP = 1/2$. Their harmonic mean is 0.4.

*Table 3: An Illustration of Cluster-F Calculation*

| Truth Clusters (T) | Predicted Clusters (P) | Calculation |
|---|---|---|
| $T_1$ = (1, 2, 3) <br> $T_2$ = (4, 5) <br> $T_3$ = (6, 7, 8) | $P_1$ = (1, 2, 3) <br> $P_2$ = (4, 5, 6, 7, 8) | cR = 1/3 = 0.3333 |
| | | cP = 1/2 = 0.5 |
| | | cF = (2×1/3×1/2)/(1/3+1/2) = 0.4 |

The calculation of $cR$ and $cP$ can be implemented as follows.

```
Algorithm 2: Cluster-F
      P, p, pIndex, T, t, tMap # same as in Algorithm 1 hereafter
  1   cSize: a hash of a cluster P_i and its size
      cMatch: the count of P_i that contains all and the only instances in T_j
  2   pIndex ← {}
  3   for each P_i ∈ P do
  4       for each p ∈ P_i do
  5           pIndex[p] ← i
  6       end for
```

```
 7        cSize[i] ← |P_i|
 8      end for
 9      cMatch ← 0
10      for each T_j ∈ T do
11        tMap ← {}
12        for each t ∈ T_j do
13          if pIndex[t] ∉ keys(tMap) then
14            tMap[pIndex[t]] ← 0
15          end if
16          tMap[pIndex[t]] ← tMap[pIndex[t]] + 1
17        end for
18        for each (key, value) ∈ tMap do
19          if value = |T_j| and cSize[key] = |T_j| then
20            cMatch ← cMatch + 1
21          end if
22        end for
23      end for
24      cR ← cMatch/|T|
25      cP ← cMatch/|P|
26      return cR, cP
```

In Algorithm 2, the code lines added to Algorithm 1 are highlighted in bold. As a result of running the skeleton code, the hash table *tMap* records every cluster index *i* associated with name instances in *T* and the frequency of each index. If (1) an index *i* (*key*)'s frequency in *tMap* is the same as the size of a truth cluster $T_j$ (*value* = $|T_j|$) and (2) the size of the cluster $P_i$ is the same (*cSize[key]* = $|T_j|$), this means that all and only name instances in the truth cluster appear together in the same predicted cluster. This is a case of the intersection ($|P \cap T|$) and increments *cMatch*, the numerator for *cR* and *cP*.

Cluster Level: K-metric

K-metric consists of Average Author Purity (*AAP*), Average Cluster Purity (*ACP*), and their geometric mean (*K*) (Santana et al., 2017).

$$AAP = \frac{1}{N} \sum_{j=1}^{|T|} \sum_{i=1}^{|P|} \frac{n_{ij}^2}{n_j} \quad (4)$$

$$ACP = \frac{1}{N} \sum_{i=1}^{|P|} \sum_{j=1}^{|T|} \frac{n_{ij}^2}{n_i} \quad (5)$$

$$K = \sqrt{ACP \times AAP} \quad (6)$$

Here, *T* and *P* represent sets of truth and predicted clusters each. *N* is the total of name instances to be disambiguated. Assume that every name instance in truth clusters is assigned to one of predicted clusters throughout this paper. $n_{ij}$ is the number of $P_i$ name instances that also appear in $T_j$; $n_i$ and $n_j$ represent the numbers of name instances in $P_i$ and $T_j$, respectively. *AAP* measures the fragmentation of truth clusters. In other words, its value is low when many instances of an *author* (= a truth cluster) are split into separate predicted clusters (≈ recall). In contrast, *ACP* measures the purity of the predicted clusters. The *ACP* value decreases if predicted clusters contain name instances that should belong to other predicted clusters (≈ precision).

Table 4 illustrates the *K-metric* calculation. *AAP* starts by counting the number of name instances in the truth cluster that appear in each predicted cluster. For example, all instances in $T_1$ appear together in $P_1$, thus $n_{11}^2 = 3^2 (= 9)$ and $n_1 = 3$. This repeats over other truth clusters ($T_2 = 2^2/2$ and $T_3 = 3^2/3$). The same procedure is applied for *ACP* but this time staring from $P_1$ being compared to each truth cluster.

*Table 4: An Illustration of K-metric Calculation*

| Truth Clusters (T) | Predicted Clusters (P) | Calculation |
|---|---|---|
| $T_1$ = (1, 2, 3) | $P_1$ = (1, 2, 3) | AAP = $(3^2/3+2^2/2+3^2/3)/8$ = 1.0 |
| $T_2$ = (4, 5) | $P_2$ = (4, 5, 6, 7, 8) | ACP = $(3^2/3+2^2/5+3^2/5)/8$ = 0.7 |
| $T_3$ = (6, 7, 8) | | K = $\sqrt{1.0 \times 0.7}$ = 0.8367 |

Equations 4 and 5 can be re-written using a set notation as follows. The order of cluster comparison (truth → predicted or predicted → truth) does not affect the calculation outcome because the final set of intersection ($P_i \cap T_j$) are the same. So, the summation can be ordered as truth clusters being compared to predicted clusters (i.e., truth → predicted) for both *AAP* and *ACP*.

$$AAP = \frac{1}{N}\sum_{j=1}^{|T|}\sum_{i=1}^{|P|}\frac{n_{ij}^2}{n_j} = \frac{1}{N}\sum_{j \in T}\sum_{i \in P}\frac{|P_i \cap T_j|^2}{|T_j|} \quad (7)$$

$$ACP = \frac{1}{N}\sum_{i=1}^{|P|}\sum_{j=1}^{|T|}\frac{n_{ij}^2}{n_i} = \frac{1}{N}\sum_{j \in T}\sum_{i \in P}\frac{|P_i \cap T_j|^2}{|P_i|} \quad (8)$$

The revised equations can be implemented expanding Algorithm 1.

```
Algorithm 3: K-metric
     P, p, pIndex, T, t, tMap
     cSize: a hash of a cluster P_i and its size
1    instSum: the total of name instances to be disambiguated
     aapSum, acpSum: totals of aap and acp values per cluster
2    pIndex ← {}
3    for each P_i ∈ P do
4        for each p ∈ P_i do
5            pIndex[p] ← i
6        end for
7        cSize[i] ← |P_i|
8    end for
9    instSum ← 0
10   for each T_j ∈ T do
11       instSum ← instSum + |T_j|
12       tMap ← {}
13       for each t ∈ T_j do
14           if pIndex[t] ∉ keys(tMap) then
15               tMap[pIndex[t]] ← 0
16           end if
17           tMap[pIndex[t]] ← tMap[pIndex[t]] + 1
```

```
18        end for
19        for each (key, value) ∈ tMap do
20            aapSum ← aapSum + value²/|Tⱼ|
21            acpSum ← acpSum + value²/|cSize[key]|
22        end for
23    end for
24    AAP ← aapSum/instSum
25    ACP ← acpSum/instSum
26    return AAP, ACP
```

Algorithm 3 recycles the skeleton code. The added lines to Algorithm 1 are shown in bold. The re-use is possible because in Equation 7 and 8, K-metric is calculated in a single procedure in which truth clusters are compared to predicted clusters for both *AAP* and *ACP*. In contrast, Equation 4 and Equation 5 formulate that truth clusters are compared to predicted clusters for *AAP* and then predicted clusters to truth clusters for *ACP*.

As all name instances in truth clusters are assigned to one of predicted clusters, the value of *N* can be obtained by counting instances in either truth (*instSum*, code line #11) or predicted clusters. In code lines #20~21, $|P_i \cap T_j|^2/|T_j|$ in Equation 7 and $|P_i \cap T_j|^2/|P_i|$ in Equation 8 are calculated and summed into *aapSum* and *acpSum*, respectively, using the hash values in *tMap*. Especially, $|P_i|$ is obtained by referencing a predicted cluster index *i* (*key*) to *cSize* generated in code line #7.

Cluster Level: Splitting & Lumping Error

Several studies have adopted the concept of Lumping (= merging) and Splitting Error (Kim & Diesner, 2016; Lerchenmueller & Sorenson, 2016; Li et al., 2014; Liu et al., 2014; Torvik & Smalheiser, 2009). Splitting Error (*SE*) and Lumping Error (*LE*) are defined as follows (Li et al., 2014):

$$SE = \frac{\sum_a |\{x | x \in T_a,\ x \notin P_a\}|}{\sum_a |T_a|} \quad (9)$$

$$LE = \frac{\sum_a |\{x | x \in P_a,\ x \notin T_a\}|}{\sum_a |P_a|} \quad (10)$$

Here, $x$ means an author name instance. $T_a$ represents the truth cluster of a unique author *a*, while $P_a$ means the predicted cluster that contains the largest number of name instances of the unique author *a*. *SE* evaluates how many name instances of a unique author (= a truth cluster) fail to appear in the predicted cluster that contains the largest number of name instances associated with the unique author (≈ recall). *LE* measures how many name instances in a predicted cluster belong to other distinct authors, i.e., truth clusters (≈ precision). Note that *SE* and *LE* consider only a predicted cluster that contains the largest number of name instances of a truth cluster. In contrast, Cluster-F considers only the perfect match of all name instances between a predicted cluster and a truth cluster. Others (K-metric, Pairwise-F, and B[3]) consider all intersection sets of instances between a truth cluster and predicted clusters.

Table 5 illustrates how to calculate *SE* and *LE*. The *SE* calculation starts by comparing name instances in T₁ with P₁ and P₂. P₁ contains the largest number of T₁ name instances. As there is no name instance in T₁ that does not belong to P₁, the value for $|\{x | x \in T_a,\ x \notin P_a\}|$ in Equation 9 is zero. Likewise, no splitting error case is detected for T₂ and T₃ because all name instances in T₂ and T₃ are found in P₂, the predicted cluster that contains all name instances of both $T_2$ and $T_3$. Thus, the numerator for *SE* is 0, while its denominator, sum of all truth cluster sizes, is 8. For *LE*, name instances in $T_1$ are all found in $P_1$. But name instances in $T_2$ and $T_3$ are lumped with those from $T_3$ and $T_2$, respectively, in the same predicted cluster $P_2$. Regarding the error for $T_2$, three name instances from $T_3$ are wrongly assigned to $P_2$ (thus, lumping error = 3), while for $T_3$, two instances from $T_2$ are wrongly assigned to $P_2$ (thus, lumping error = 2). As both $T_2$ and $T_3$ share the largest predicted cluster, $P_2$, their $|P_a|$ value is 5 (=$|P_2|$).

*Table 5: An Illustration of Splitting & Lumping Errors Calculation*

| Truth Clusters (T) | Predicted Clusters (P) | Calculation |
|---|---|---|
| $T_1$ = (1, 2, 3) $T_2$ = (4, 5) $T_3$ = (6, 7, 8) | $P_1$ = (1, 2, 3) $P_2$ = (4, 5, 6, 7, 8) | SE = (0+0+0)/(3+2+3) = 0.0 |
| | | LE = (0+3+2)/(3+5+5) = 0.3846 |

A key difference between *SE* & *LE* and other four measures is that *SE* & *LE* counts errors (split or lumped name instances), while others count correctly predicted name instances. For the comparison across five measures, these error-based measures can be converted into recall (*eR*), precision (*eP*), and F (*eF*) measures as follows (Lerchenmueller & Sorenson, 2016; Liu et al., 2014; Torvik & Smalheiser, 2009):

$$eR = 1 - SE \quad (11)$$

$$eP = 1 - LE \quad (12)$$

$$eF = \frac{2 \times eR \times eP}{eR + eP} \quad (13)$$

This conversion scales *eR* between 0 (all split) and 1 (no splitting), and *eP* between 0 (all lumped) and 1 (no lumping). In Table 5, for example, *eR* = 1 – *SE* = 1 – 0 = 1 and *eP* = 1 – *LE* = 1 – 0.3846 = 0.6154. Their harmonic mean (= 0.7619) is *eF*.

Equation 9 and 10 can be re-written using a set notation as follows.

$$SE = \frac{\sum_a |\{x | x \in T_a, \; x \notin P_a\}|}{\sum_a |T_a|} = \frac{\sum_a (|T_a| - |T_a \cap P_a|)}{\sum_a |T_a|} \quad (14)$$

$$LE = \frac{\sum_a |\{x | x \in P_a, \; x \notin T_a\}|}{\sum_a |P_a|} = \frac{\sum_a (|P_a| - |T_a \cap P_a|)}{\sum_a |P_a|} \quad (15)$$

The calculation of *SE* and *LE* can be implemented by adding lines to the skeleton code as follows.

| Algorithm 4: SE & LE | |
|---|---|
| 1 | *P, p, pIndex, T, t, tMap* **cSize: a hash of a cluster $P_i$ and its size** **spSum: sum of split instances** **lmSum: sum of lumped instances** |

```
        instTrSum: sum of instances in the truth clusters for a unique author
        instPrSum; sum of instances in the largest predicted clusters for a unique author
2       pIndex ← {}
3       for each P_i ∈ P do
4           for each p ∈ P_i do
5               pIndex[p] ← i
6           end for
7           cSize[i] ← |P_i|
8       end for
9       for each T_j ∈ T do
10          tMap ← {}
11          for each t ∈ T_j do
12              if pIndex[t] ∉ keys(tMap) then
13                  tMap[pIndex[t]] ← 0
14              end if
15              tMap[pIndex[t]] ← tMap[pIndex[t]] + 1
16          end for
17          maxKey ← 0, maxValue ← 0
18          for each (key, value) ∈ indexMap do
19              if value > maxValue then
20                  maxValue ← value
21                  maxKey ← key
22              end if
23          end for
24          spSum ← spSum + (|T_j| − maxValue)
25          lmSum ← lmSum + (cSize[maxKey] − maxValue)
26          instTrSum ← instTrSum + |T_j|
27          instPrSum ← instPrSum + cSize[maxKey]
28      end for
29      SE ← spSum/instTrSum
30      LE ← lmSum/instPrSum
31      return SE, LE
```

In Algorithm 4, code lines #17 and #19~#22 find the predicted cluster index *i* (*key*) with the largest frequency (*value*) from *tMap*. For an author *a* (= a truth cluster $|T_a|$), the *maxValue* in *tMap* is used for counting $|T_a \cap P_a|$ in Equation 14 and 15. In addition, the *key* for the *maxValue* is used to obtain the value for $cSize[maxKey] = |P_a|$, which is the size of the predicted cluster that contains the largest number of name instances in the truth cluster $|T_a|$.

Pairwise Level: Pairwise-F

This measures disambiguation performance at a pair-level via pairwise Precision (*pP*), pairwise Recall (*pR*), and Pairwise-F1 (*pF*) as defined below (Menestrina et al., 2010):

$$pR = \frac{|pairs(P) \cap pairs(T)|}{|pairs(T)|} \quad (16)$$

$$pP = \frac{|pairs(P) \cap pairs(T)|}{|pairs(P)|} \quad (17)$$

$$pF = \frac{2 \times pR \times pP}{pR + pP} \quad (18)$$

Here, *pairs(P)* and *pairs(T)* mean name instance pairs generated from the same cluster in predicted clusters P and truth clusters T. The numerator $|pairs(P) \cap pairs(T)|$ is the number of instance pairs that appear both in P and T.

The calculation of *pR* and *pP* is illustrated in Table 6. Here, a pair of name instances is represented by two instance ids separated by a vertical bar. In T1, for example, three name instances (1, 2, and 3) are paired into three pairs (1|2, 1|3, and 2|3). The list of name pairs of truth clusters is compared with that of predicted clusters to generate a list of pairs found in both lists. The count of these intersection pairs constitutes the numerator (1|2, 1|3, 2|3, 4|5, 6|7, 6|8, 7|8; 7 pairs), which is divided by the total of pairs in truth clusters (= 7) for *pR* and by the total of pairs in predicted clusters (=13) for *pP*.

*Table 6: An Illustration of Pairwise-F Calculation*

| Truth Clusters (T) | Predicted Clusters (P) | Calculation |
| --- | --- | --- |
| $T_1$ = (1, 2, 3) → (1|2, 1|3, 2|3) | $P_1$ = (1, 2, 3) → (1|2, 1|3, 2|3) | pR = 7/7 = 1.0 |
| $T_2$ = (4, 5) → (4|5) | $P_2$ = (4, 5, 6, 7, 8) → (4|5, 4|6, 4|7, 4|8, 5|6, 5|7, 5|8, 6|7, 6|8, 7|8) | pP = 7/13 = 0.5385 |
| $T_3$ = (6, 7, 8) → (6|7, 6|8, 7|8) | | pF = 2×(1.0×0.5385)/(1.0+0.5385) = 0.7000 |

Calculating *pR* and *pP* can be memory- and time-consuming because the number of pairs in a cluster increases in a quadratic way with the size of name instances (Levin et al., 2012; Louppe, Al-Natsheh, Susik, & Maguire, 2016). For example, the number of pairs for a cluster with 10 instances is 45, while that of a cluster with 1,000 instances is 499,500. To overcome this problem, the *Pairwise-F* measures can be re-written as follows.

$$pR = \frac{|pairs(P) \cap pairs(T)|}{|pairs(T)|} = \frac{\sum_{j \in T} \sum_{i \in P} |T_j \cap P_i| \times (|T_j \cap P_i| - 1)/2}{\sum_{j \in T} |T_j| \times (|T_j| - 1)/2} \quad (19)$$

$$pP = \frac{|pairs(P) \cap pairs(T)|}{|pairs(P)|} = \frac{\sum_{j \in T} \sum_{i \in P} |T_j \cap P_i| \times (|T_j \cap P_i| - 1)/2}{\sum_{i \in P} |P_i| \times (|P_i| - 1)/2} \quad (20)$$

Here, the number of pairs in a cluster is counted not by generating all possible pairs in the cluster but by a heuristic that the number of pairs in a cluster can be calculated from the number of instances in a cluster via cluster size × (cluster size – 1)/2. Likewise, the number of pairs in an intersection can be obtained from the number of instances in it. Algorithm 4 implements this heuristic.

```
Algorithm 5: Pairwise-F
  P, p, pIndex, T, t, tMap
  pairPrSum: the total of instance pairs in predicted clusters
1 pairTrSum: the total of instance pairs in truth clusters
  pairIntSum: the total of instance pairs in the intersection of predicted and truth clusters
2 pIndex ← {}
3 for each P_i ∈ P do
4     for each p ∈ P_i do
5         pIndex[p] ← i
```

```
6           end for
7           pairPrSum ← pairPrSum + |P_i| × (|P_i| − 1)/2
8       end for
9       for each T_j ∈ T do
10          pairTrSum ← pairTrSum + |T_j| × (|T_j| − 1)/2
11          tMap ← {}
12          for each t ∈ T_j do
13              if pIndex[t] ∉ keys(tMap) then
14                  tMap[pIndex[t]] ← 0
15              end if
16              tMap[pIndex[t]] ← tMap[pIndex[t]] + 1
17          end for
18          for each (key, value) ∈ indexMap do
19              pairIntSum ← pairIntSum + |value| × (|value| − 1)/2
20          end for
21      end for
22      pR ← pairIntSum/pairTrSum
23      pP ← pairIntSum/pairPrSum
24      return pR, pP
```

Again, this implementation of *pR* and *pP* is based on the same skeleton code for K-metric and SE & LE as well as Cluster-F. The added code to Algorithm 1 are highlighted in bold.

Instance Level: B-Cubed

This measures clustering performance at an instance-level. Three parts of this measure – $B^3$ Recall (*bR*), $B^3$ Precision (*bP*), and $B^3$ F (*bF*) – are defined as follows (Levin et al., 2012):

$$bR = \frac{1}{N} \sum_{t \in T} \frac{|P(t) \cap T(t)|}{|T(t)|} \qquad (21)$$

$$bP = \frac{1}{N} \sum_{t \in T} \frac{|P(t) \cap T(t)|}{|P(t)|} \qquad (22)$$

$$bF = \frac{2 \times bR \times bP}{bR + bP} \qquad (23)$$

Here, *t* is a name instance in truth clusters *T*. *N* is the number of all name instances in truth clusters (*T*). $T(t)$ means a truth cluster that contains a name instance *t*, while $P(t)$ means a predicted cluster that contains the name instance *t*.

Table 7 shows an illustration of $B^3$ calculation. Starting with the instance 1 in $T_1$ for *bR*, for example, a predicted cluster containing it is detected: $P(1) = P_1$ and $(1) = T_1$ . Next, the intersection of the truth cluster ($T_1$) and the predicted cluster ($P_1$) is filtered (1, 2, and 3). Then, $|P_1 \cap T_1|/|T_1|$ = 3/3 is obtained. This is repeated for instances 2 and 3 in $T_1$, resulting in an array of (3/3, 3/3, 3/3) for $T_1$. After the same procedure is applied to $T_2$ and $T_3$, the sum of $|P(t) \cap T(t)|/|T(t)|$ for all name instances is divided by the total of those instances (= 8), producing *bR* = 1.0.

*Table 7: An Illustration of B³ F Calculation*

| Truth Clusters (T) | Predicted Clusters (P) | Calculation |
|---|---|---|
| $T_1 = (1, 2, 3)$<br>$T_2 = (4, 5)$<br>$T_3 = (6, 7, 8)$ | $P_1 = (1, 2, 3)$<br>$P_2 = (4, 5, 6, 7, 8)$ | bR = ((3/3+3/3+3/3)+(2/2+2/2)+(3/3+3/3+3/3))/8 = 1.0 |
| | | bP = ((3/3+3/3+3/3)+(2/5+2/5+3/5+3/5+3/5))/8 = 0.7 |
| | | bF = 2×(1.0×0.7)/(1.0+0.7) = 0.8235 |

Although B³ is an instance level metric, its calculation can be formulated as a cluster-level calculation. This is possible because in Equation 21 and 22, the calculation results for each name instance in the same intersection are the same. In Table 7, for example, instances 4 and 5 in $T_2$ have the same calculation outcome (= 2/2) as they appear together in the intersection of $T_2$ and $P_2$. So, we can re-write (2/2 + 2/2) as (2/2)×2 = $2^2$/2. Here, 2/2 (or $2^2$) is the calculation outcome for an instance and 2 besides 2/2 is the number of instances in the intersection ($|T_2 \cap P_2|$). Drawing on this formulation, Equation 21 and 22 can be re-written as follows.

$$bR = \frac{1}{N}\sum_{t \in T}\frac{|P(t) \cap T(t)|}{|T(t)|} = \frac{1}{N}\sum_{j \in T}\sum_{t \in T_j}\frac{|P(t) \cap T_j|}{|T_j|} = \frac{1}{N}\sum_{j \in T}\sum_{t \in T_j}\sum_{i \in P}\frac{|P_i \cap T_j|}{|T_j|}$$

$$= \frac{1}{N}\sum_{j \in T}\sum_{i \in P}\frac{|P_i \cap T_j|}{|T_j|} \times |P_i \cap T_j| = \frac{1}{N}\sum_{j \in T}\sum_{i \in P}\frac{|P_i \cap T_j|^2}{|T_j|} = AAP \quad (24)$$

$$bP = \frac{1}{N}\sum_{t \in T}\frac{|P(t) \cap T(t)|}{|P(t)|} = \frac{1}{N}\sum_{j \in T}\sum_{t \in T_j}\frac{|P(t) \cap T_j|}{|P(t)|} = \frac{1}{N}\sum_{j \in T}\sum_{t \in T_j}\sum_{i \in P}\frac{|P_i \cap T_j|}{|P_i|}$$

$$= \frac{1}{N}\sum_{j \in T}\sum_{i \in P}\frac{|P_i \cap T_j|}{|P_i|} \times |P_i \cap T_j| = \frac{1}{N}\sum_{j \in T}\sum_{i \in P}\frac{|P_i \cap T_j|^2}{|P_i|} = ACP \quad (25)$$

In Equation 24, a cluster $T_j$ is set first as a calculation unit ($\sum_{j \in T}\sum_{t \in T_j}()$). This follows the transformation of $T(t)$ to $T_j$ because all name instances in $T_j$ have the same set elements (themselves) and thus the same value for $|T(t)|$ ($= |T_j|$). Next, an instance $t$ needs to be checked cluster by cluster to decide where it appears in predicted clusters $P_i(t)$ as in $\sum_{j \in T}\sum_{t \in T_j}\sum_{i \in P}|P_i(t) \cap T_j|/|T_j|$. Evidently, $P_i(t)$ is the same as $P_i$. Finally, the calculation process can be simplified as $\sum_{j \in T}\sum_{i \in P}|P_i \cap T_j|/|T_j| \times |P_i \cap T_j|$. This is because the calculation results of $|P_i \cap T_j|/|T_j|$ for name instances in the same cluster are the same if the instances appear in the same intersection ($P_i \cap T_j$). That is why $|P_i \cap T_j|/|T_j|$ is multiplied by the number of instances belonging to the intersection ($|P_i \cap T_j|$), omitting the part of instance referencing in the nested summation ($\sum_{t \in T_j}()$). The final re-writing is the same as the calculation of $AAP$ in Equation 7. Likewise, $bP$ can be re-written to match $ACP$ (Equation 25). This transformation can be illustrated by the example in Table 8, where the calculation for B³ and K-metric is juxtaposed to show their similarity.

Table 8: An Illustration of $B^3$ F Calculation in comparison with K-metric Calculation

| Truth Clusters (T) | Predicted Clusters (P) | Calculation |
|---|---|---|
| $T_1$ = (1, 2, 3)<br>$T_2$ = (4, 5)<br>$T_3$ = (6, 7, 8) | $P_1$ = (1, 2, 3)<br>$P_2$ = (4, 5, 6, 7, 8) | bR = ((3/3+3/3+3/3)+(2/2+2/2)+(3/3+3/3+3/3))/8 = 1.0<br>AAP = (($3^2$/3)+($2^2$/2)+($3^2$/3))/8 = 1.0<br>bP = ((3/3+3/3+3/3)+(2/5+2/5+3/5+3/5+3/5))/8 = 0.7<br>ACP = (($3^2$/3)+($2^2$/5+$3^2$/5))/8 = 0.7<br>bF = 2×(1.0×0.7)/(1.0+0.7) = 0.8235<br>K = $\sqrt{1.0 \times 0.7}$ = 0.8367 |

As such, Equations 24 and 25 indicate that bR and bP can be calculated by Algorithm 3 for calculating AAP and ACP. A difference is that $B^3$ F is a harmonic mean of AAP (= bR) and ACP (= bP), while K is a geometric mean of AAP and ACP.

All-in-one Calculation and Runtime Test

In Algorithm 2 ~ 5, the five clustering measures have been shown to be calculable using the same skeleton code in Algorithm 1. This commonality enables us to integrate those five measures in a single set of code, as in Algorithm 6 below. Note that $B^3$ precision and recall are not calculated because they are the same as ACP and AAP in K-metric.

```
Algorithm 6: All-In-One
     P, p, pIndex, T, t, tMap #common to all measures
     cSize #Cluster-F, K-metric, B³, SE&LE
     cMatch #Cluster-F
1    instSum: #K-metric, B³
     aapSum, acpSum #K-metric, B³
     spSum, lmSum, instTrSum, instPrSum # SE&LE
     pairPrSum, parSumTr, pairIntSum #Pairwise-F
2    pIndex ← {}
3    for each P_i ∈ P do
4        for each p ∈ P_i do
5            pIndex[p] ← i
6        end for
7        cSize[i] ← |P_i|
8        pairPrSum ← pairPrSum + |P_i| × (|P_i| − 1)/2
9    end for
10   instSum ← 0
11   for each T_j ∈ T do
12       instSum ← instSum + |T_j|
13       pairTrSum ← pairTrSum + |T_j| × (|T_j| − 1)/2
14       tMap ← {}
15       for each t ∈ T_j do
16           if pIndex[t] ∉ keys(tMap) then
17               tMap[pIndex[t]] ← 0
18           end if
19           tMap[pIndex[t]] ← tMap[pIndex[t]] + 1
20       end for
21       maxKey ← 0, maxValue ← 0
22       for each (key, value) ∈ indexMap do
23           if value = |T_j| and cSize[key] = |T_j| then
24               cMatch ← cMatch + 1
25           end if
26           aapSum ← aapSum + value²/|T_j|
```

```
27        acpSum ← acpSum + value²/|cSize[key]|
28        if value > maxValue then
29            maxValue ← value
30            maxKey ← key
31        end if
32        pairIntSum ← pairIntSum + |value| × (|value| − 1)/2
33      end for
34      spSum ← spSum + (|T_j| − maxValue)
35      lmSum ← lmSum + (cSize[maxKey] − maxValue)
36      instTrSum ← instTrSum + |T_j|
37      instPrSum ← instPrSum + cSize[maxKey]
38   end for
39   cR ← cMatch/|T|, cP ← cMatch/|P|
40   AAP ← aapSum/instSum, ACP ← acpSum/instSum
41   SE ← spSum/instTrSum, LE ← lmSum/instPrSum
42   pR ← pairIntSum/pairTrSum, pP ← pairIntSum/pairPrSum
43   return cR, cP, AAP (bR), ACP (bP), SE, LE, pR, pP
```

Besides integrating multiple measures in a single framework, the algorithm greatly reduces computation time. To illustrate this, a total of 41,358 name instances in KISTI were used to evaluate the clustering performance of DBLP's disambiguation by the five measures as in Figure 2. For this, especially, the steps implied in the original equations of the five measures were implemented straightforwardly. For example, instance pairs per cluster were generated (797,297 truth pairs and 826,187 predicted pairs) and compared for intersection one by one. Execution time of each measure was measured in seconds and compared to that of the same measure implemented by its corresponding Algorithm 2~5[4]. Table 9 reports the runtime results.

*Table 9: Runtime (in Seconds) of Measures Implemented by Straightforward vs Proposed Algorithms*

| Calculation | Cluster-F | K-metric | SE & LE | Pairwise-F | B³ |
|---|---|---|---|---|---|
| Straightforward | 46.920 | 231.975 | 119.925 | 23433.140 | 138.956 |
| Proposed (Algorithm 2~5) | 0.025 | 0.055 | 0.057 | 0.055 | 0.055 |
| All-In-One (Algorithm 6) | 0.064 | | | | |

Table 9 reports that Algorithm 2~5 calculated each measure less than 0.057 seconds, while the straightforward implementations took approximately 47 (Cluster-F) up to 23,433 (6.5 hour, Pairwise-F) seconds. All measures could be calculated in less than 0.065 seconds by the All-In-One algorithm.

To test the scalability of Algorithm 6, a set of 1.2 M name instances associated with unique identifiers in a high-energy physics publication library, INSPIRE, was obtained (Louppe et al., 2016). Using the INSPIRE unique identifiers as the ground truth of author identity, the performance of all-initials-based name disambiguation[5] was evaluated by the five measures. This task is challenging, especially for the calculation of Pairwise-F, because the number of instance pairs in truth clusters (= 15,388 authors) approximates 213.4 M, while that in predicted clusters (= 18,672) was almost 194.5 M (intersection pairs

---

[4] Runtime was tested on a desktop with Intel Core i7-7700 CPU (3.60GHz), 32G RAM, and 64-bit Windows OS by running code in Strawberry Perl 64-bit (ver. 5.26). Runtime was tested 10 times for each measure and the best result was reported for each.
[5] Two name instances that share the same full surname and initials of all forenames are predicted to refer to the same person. For details, see Kim (2018).

≈ 179.9 M). Algorithm 6 produced evaluation results by all five measures in 1.583 seconds. Tested only for the Pairwise-F calculation by Algorithm 5, the runtime was 1.552 seconds, which is comparable to 12.903 seconds by the Generalized Merged Distance (GMD) algorithm[6] (Menestrina et al., 2010), the most runtime-efficient method for calculating Pairwise-F so far.

## Conclusion and Discussion

This paper demonstrated that five measures of clustering performance in author name disambiguation can be integrated into one calculation framework. This was possible mainly because name instances in truth and predicted clusters were compared not by a brute-force cluster-by-cluster comparison but by the use of two hash tables recording instances with their predicted cluster indices and their frequencies in the predicted-truth cluster intersection. Using set notations, each measure's equations was formulated to fit into the integrative framework.

A few contributions of this paper are worth noting. First, as there is no standard collection of code for the five performance measures above, this paper can provide an anchoring place for scholars to implement them and validate their correctness with efficient code and samples. Second, the measurement integration dramatically reduces runtime compared to the straightforward implementation of those measures mainly due to the use of hash tables instead of brute-force cluster-by-cluster and instance-by-instance comparisons that can increase runtime up to $O(n^2)$. Especially, Pairwise-F was re-formulated using a heuristic for counting pairs in a cluster. The scalability of the integrative calculation can help scholars evaluate the clustering performance of a disambiguation method at a large scale, for example, using several millions of name instances associated with Researcher IDs in Web of Science (Backes, 2018). This paper demonstrated this potential by evaluating the clustering results of 1.2M name instances.

Another contribution is that K-metric and $B^3$ measures were shown to produce the same recall and precision scores. This means that studies using either K-metric or $B^3$ have evaluated their clustering results by the basically same measures and thus are directly comparable to one another. Also, this can be good news to scholars who use K-metric because $B^3$ has been argued to evaluate clustering results better than others on challenging cases (Amigó et al., 2009). In addition, the usage frequency of these two different-but-same measures in Table 1 equals that of Pairwise-F (= 15), which makes them a family of major measurement in author name disambiguation.

Most importantly, the integrative calculation shows that the five measures for clustering performance in author name disambiguation can be understood within a single framework for their similarities and differences. This can help us modify current measures or propose new measures that assess disambiguation performance from distinctive perspectives. In addition, this integrative framework can incorporate other clustering measures such as Closest-Cluster-F (Menestrina et al., 2010) and Variation of Information (Meilă, 2003), which have been rarely used in author name disambiguation. Such integration will not only guide us to select measures characterizing best disambiguation performance but also help future efforts to compare different evaluation schemes under diverse ambiguity conditions for entity resolution in general beyond author name disambiguation.

## Acknowledgements

This work was supported by grants from the National Science Foundation (#1561687 and #1535370), the Alfred P. Sloan Foundation and the Ewing Marion Kauffman Foundation.

---

[6] The GMD method was implemented by Algorithm 1 in Menestrina et al. (2010).